\documentclass[aps,pre,reprint,superscriptaddress]{revtex4-2}

\usepackage[utf8]{inputenc}

\usepackage{amsmath}
\usepackage{amssymb}
\usepackage{bm}

\usepackage{hyperref}

\usepackage{graphicx}

\usepackage{listings}

\begin{document}
	
	
	\title{Pseudo-laminar chaos from on-off intermittency}
	
	
	\author{David Müller-Bender}
	\email[]{david.mueller-bender@mailbox.org}
	\affiliation{Institute of Physics, Chemnitz University of Technology, 09107 Chemnitz, Germany}
	\author{Rahil N. Valani}
	\email[]{rahil.valani@adelaide.edu.au}
	\affiliation{School of Mathematical Sciences, University of Adelaide, Adelaide, South Australia 5005, Australia}
	\author{Günter Radons}
	\email[]{radons@physik.tu-chemnitz.de}
	\affiliation{Institute of Physics, Chemnitz University of Technology, 09107 Chemnitz, Germany}
	\affiliation{ICM - Institute for Mechanical and Industrial Engineering, 09117 Chemnitz, Germany}
	
	
	\date{\today}
	
	\begin{abstract}
		In finite-dimensional, chaotic, Lorenz-like wave-particle dynamical systems one can find diffusive trajectories, which share their appearance with that of laminar chaotic diffusion [Phys. Rev. Lett. 128, 074101 (2022)] known from delay systems with lag-time modulation.
		Applying, however, to such systems a test for laminar chaos, as proposed in [Phys. Rev. E 101, 032213 (2020)], these signals fail such test, thus leading to the notion of pseudo-laminar chaos.
		The latter can be interpreted as integrated periodically driven on-off intermittency.
		We demonstrate that, on a signal level, true laminar and pseudo-laminar chaos are hardly distinguishable in systems with and without dynamical noise. However, very pronounced differences become apparent when correlations of signals and increments are considered.
		We compare and contrast these properties of pseudo-laminar chaos with true laminar chaos.
	\end{abstract}
	
	
	\maketitle
	
	\section{Introduction}
	
	Recently, the phenomenon of laminar chaos was detected in systems with periodically modulated delay \cite{muller_laminar_2018}.
	It is characterized by nearly constant laminar phases, which are periodically interrupted by irregular bursts, where the intensity levels of the laminar phases vary chaotically. 
	In contrast to its generalizations \cite{muller-bender_resonant_2019}, laminar chaotic signals with their plateau-like structure have potential advantages in applications, such as random number generators \cite{uchida_fast_2008,reidler_ultrahigh-speed_2009,kanter_optical_2010}, chaos communication \cite{goedgebuer_optical_1998,vanwiggeren_optical_1998,udaltsov_communicating_2001,keuninckx_encryption_2017}, or reservoir computing \cite{appeltant_information_2011,larger_photonic_2012,paquot_optoelectronic_2012,larger_high-speed_2017,hart_delayed_2019}, also because they can be generated experimentally e.g. by optoelectronic \cite{hart_laminar_2019} or electronic setups \cite{jungling_laminar_2020,kulminskii_laminar_2020}.
	The appealing feature of such systems is the possibility to generate signals, which are basically piecewise constant, while the levels of their intensity varies chaotically as ruled by a deterministic one-dimensional iterated map, which is robust against noise \cite{hart_laminar_2019,muller-bender_laminar_2020} and can be designed by tuning the delayed nonlinearity of the system, for instance, via electronic circuits \cite{banerjee_time-delayed_2018}.
	The dynamics of the durations of the laminar phases is governed by the modulated delay, which can be experimentally implemented and tuned by a variety of methods \cite{hart_laminar_2019,jungling_laminar_2020,kulminskii_laminar_2020,karmakar_oscillating_2020}.
	Very recently we found that for a large class of nonlinearities such systems may also show deterministic chaos with a diffusive variation of the intensity levels \cite{albers_chaotic_2022}.
	In this paper we will show that finite-dimensional ordinary differential equations (ODEs) may show a similar piecewise constant, diffusive variation of the intensity, which poses the question whether their behavior may also be considered as an instance of laminar chaos.
	The specific system of ODEs to be considered is derived from the dynamics of a one-dimensional classical wave-particle entity~\citep{phdthesismolacek,Durey2020lorenz,rahman_walking_2020,ValaniUnsteady,Valanilorenz2022}. Such wave-particle entities are motivated from the hydrodynamical system of walking droplets and they have been shown to mimic features that are typically associated with quantum mechanical systems~\cite{bush_hydrodynamic_2020}. The corresponding evolution equations considered by us, are modifications of the classic Lorenz equations \cite{lorenz_deterministic_1963,sparrow_lorenz_1982}.
	In \cite{muller-bender_laminar_2020}, we designed a toolbox of testing experimental time-series for laminar chaos. We now apply these methods to the numerically generated time series of the Lorenz-like wave-particle system and find that they do not pass this test.
	Therefore, to characterize their behavior and to distinguish it from “true” laminar chaos as it appears in differential delay equations (DDE), we coin the term pseudo-laminar chaos.
	It turns out that pseudo-laminar chaos can be considered as integrated on-off intermittency, more specifically, as the integrated version of periodically driven on-off intermittency.
	On-off intermittency is a fundamental phenomenon \cite{platt_on-off_1993,heagy_characterization_1994}, which is closely related to synchronization phenomena in coupled systems \cite{fujisaka_new_1985,pikovsky_symmetry_1991,boccaletti_synchronization_2002}.
	In past decades, it has been observed experimentally in many systems \cite{hammer_experimental_1994,rodelsperger_on-off_1995,john_on-off_1999,cabrera_on-off_2002}.
	In these classical papers, autonomous or randomly driven (\cite{yang_on-off_1994,yang_on-off_1996}) dissipative systems were considered, which under suitable conditions can show power-law distributed off-periods of the variable of interest.
	The power-law distribution of the off-periods can lead to anomalous transport in diffusive systems driven by on-off intermittency \cite{harada_on-off_1999,miyazaki_continuous-time_2001,sato_anomalous_2019}.
	However, the basic on-off mechanism, namely the repeated variation of a time-dependent parameter (or a second variable) through a bifurcation point \cite{heagy_characterization_1994}, can also occur with a periodic variation, leading to a trivial distribution of the off-periods.
	It is clear that the integrated version of the periodic on-off process, which leads to pseudo-laminar chaos, can also be observed frequently.
	Therefore, it is rewarding to work out in some detail similarities and differences between true laminar and pseudo-laminar chaos, as is done in this manuscript.
	
	This paper is organized as follows.
	In Sec.~\ref{sec:def} we introduce the systems of interest and show numerical solutions of their evolution equations, which are subsequently classified as laminar and pseudo-laminar chaos, respectively.
	In Sec.~\ref{sec:lamchaostests} we show the results of applying the test for laminar chaos and also investigate statistical properties, and in Sec.~\ref{sec:noise} we check the sensitivity to noise.
	We discuss and conclude in Sec.~\ref{sec:discussion}.

	\section{System definition and time series}
	\label{sec:def}

To explore the notion of pseudo-laminar chaos, we consider a finite-dimensional Lorenz-like dynamical system that is motivated from the physical system of walking droplets.

On vertically vibrating a bath of liquid, a droplet of the same liquid can be made to walk horizontally on its free surface~\citep{Couder2005,Couder2005WalkingDroplets,superwalker}. Each bounce of the droplet generates a localized standing wave on the liquid surface that decays slowly in time. The droplet then interacts with its self-generated waves on subsequent bounces to propel itself horizontally. This gives rise to a classical millimeter-sized wave-particle entity moving steadily across the liquid surface. Intriguingly, these self-propelled classical wave-particle entities have been shown to exhibit features that are typically associated with the quantum realm~\citep{Bush2015,bush_hydrodynamic_2020}.

Exploration of walking-droplet inspired wave-particle entities in a generalized pilot-wave framework have shown a plethora of rich dynamical walking behaviors. The generalized pilot-wave framework, introduced by \citet{Bush2015}, is a theoretical framework that extends models of walking droplets beyond experimentally achievable regimes. It has allowed for the exploration of a broader class of dynamical systems and the discovery of new quantum analogs~\citep{bush_hydrodynamic_2020}. \citet{Oza2013} developed a theoretical stroboscopic model that averages over the vertical periodic bouncing motion of the droplet and provides a trajectory equation for the horizontal walking dynamics in two dimensions. A one-dimensional reduction of this model describing the dynamics of an idealized one-dimensional wave-particle entity with a sinusoidal wave field can be mapped directly to the classic Lorenz system~\citep{lorenz_deterministic_1963}. This results in the following Lorenz-like dynamical system governing the particle's dynamics (in dimensionless form)~\citep{Durey2020lorenz,ValaniUnsteady,Valanilorenz2022}:
	\begin{subequations}
		\begin{align}
			\dot{x}(t) &= X(t) \label{eq:lorenzx}\\
			\dot{X}(t) &= \sigma\, [Y(t) - X(t)] \label{eq:lorenzX}\\
			\dot{Y}(t) &= X(t)\, [r - Z(t)] - Y(t) \label{eq:lorenzY}\\
			\dot{Z}(t) &= X(t)\,Y(t) - b\,Z(t) \label{eq:lorenzZ},
		\end{align}
		\label{eq:lorenz}
	\end{subequations}
where $x(t)$ is the particle position, $X(t)$ is the particle velocity and the variable $Y(t)$ and $Z(t)$ are related to the wave forcing on the particle arising from its self-generated wave field. The parameter $b$ turns out to be unity in the Lorenz-like dynamical equations of the wave-particle entity, while $\sigma^{-1}$ represents the dimensionless particle/droplet mass and $r$ represents the wave force coefficient. We refer the interested reader to \citet{ValaniUnsteady} and \citet{Valanilorenz2022} for a derivation and more details on this model.

	\begin{figure}
	\includegraphics[width=0.48\textwidth]{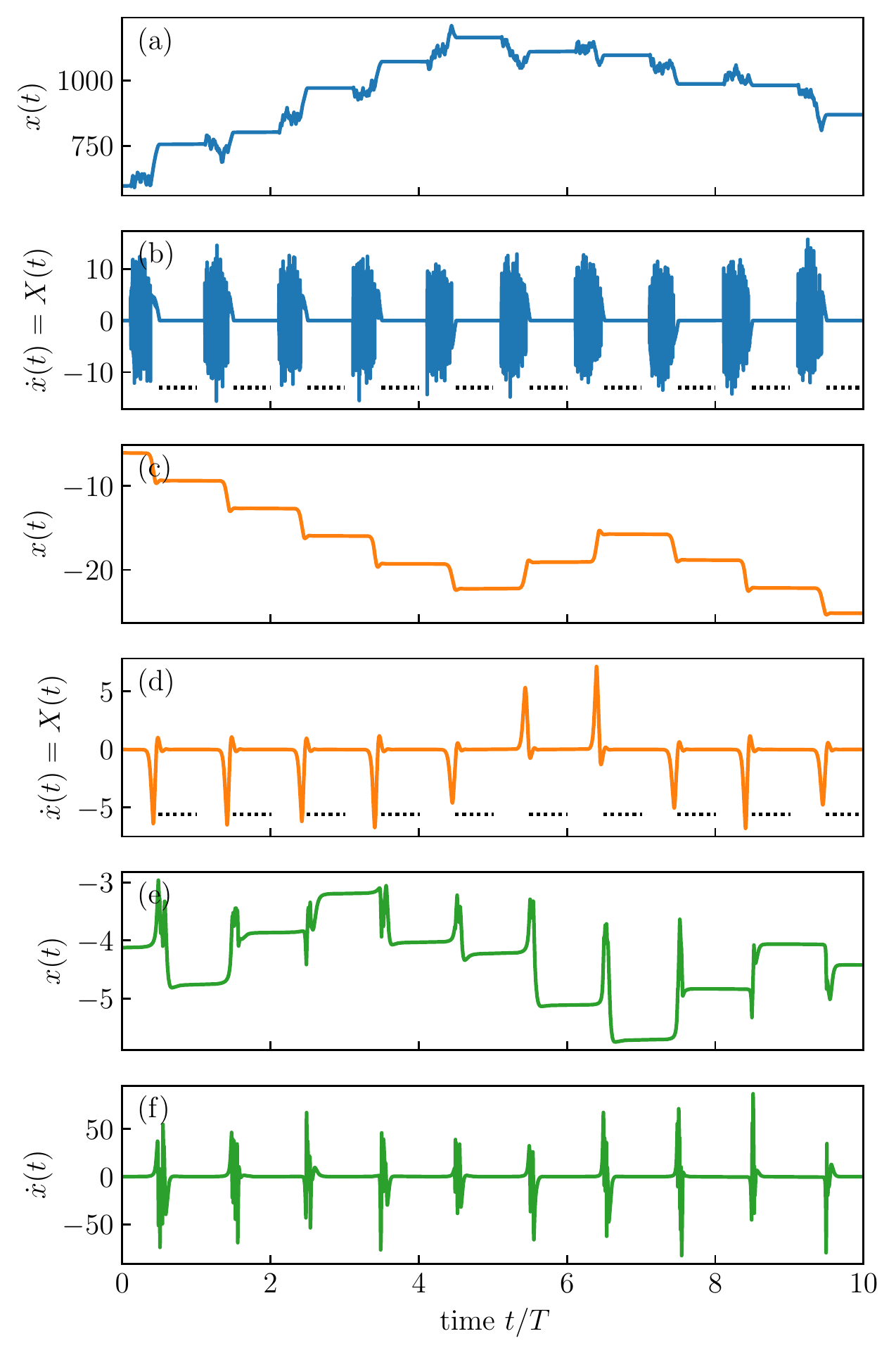}
	\caption{Time series of pseudo-laminar chaos and laminar chaos.
		In (a) and (c) pseudo-laminar chaos generated from Eq.~\eqref{eq:lorenz} is shown.
		True laminar chaos generated from Eq.~\eqref{eq:dde} is shown in (e).
		The corresponding derivatives of the time series are plotted below in Figs. (b),(d), and (f).
		For (a) and (b) we used the parameters $\sigma=5$, $b=1$, $A_r = 40$, and $T=400$.
		In (c) and (d) we have $\sigma=5$, $b=1$, $A_r = 15$, and $T=10$.
		For the laminar chaos in (e) and (f), we used $\Theta=50$, $\mu = 0.9$, $\tau_0=1$, $A_\tau = 0.76$, and $T=1$.
		For a sufficient relaxation of the transient dynamics, plotting of the time series begins after $100$ periods of the time-varying parameter $r(t)$ or of the delay $\tau(t)$.
		The dotted lines in (b) and (d) correspond to $r(t)<1$, where a stable fixed point exists at $(X,Y,Z)=(0,0,0)$ of the system in Eqs.~\eqref{eq:lorenz}.
	}
	\label{fig:traj}
	\end{figure}

Superwalkers~\citep{superwalker, superwalkernumerical} are bigger and faster walking droplets that emerge when the bath is vibrated at two frequencies simultaneously, a frequency and half of it, along with a constant phase difference. By detuning the two driving frequencies by a small amount, one can get the phase difference to drift slowly in time. This detuned two-frequency driving results in a novel walking motion for superwalkers known as stop-and-go motion~\citep{superwalker,ValaniSGM}. The stop-and-go motion of droplets, enabled by the varying phase difference, results in periodic traversals of the bouncing (no horizontal walking) and the walking regime in the parameter-space of the physical system. Hence, the stop-and-go motion of superwalkers motivates periodically varying parameters in our idealized Lorenz-like wave-particle model, Eqs.~\eqref{eq:lorenz}, which would allow us to periodically traverse between a stationary state and walking states. 

We consider a simple extension to the system in Eqs.~\eqref{eq:lorenz} by replacing the constant $r$ in Eq.~\eqref{eq:lorenzY} by the time-varying parameter $r(t) = A_r \sin\left(\frac{2\pi}{T}\, t \right)$ resulting in a parametrically driven Lorenz-like system~\citep{Han2017}~\footnote{We note that in the physical model of the wave-particle entity, we require the parameter $r(t)$ to be positive whereas the chosen form of $r(t)$ allows negative values. However, to realize pseudo-laminar chaos while keeping $r(t)>0$, we would require sophisticated periodic functions for $r(t)$, whereas the present simple sinusoidal variation allows us to explore pseudo-laminar chaos readily in the parametrically driven Lorenz-like system.}.
	The amplitude $A_r$ and the period $T$ of $r(t)$ are chosen such that Eqs.~(\ref{eq:lorenzX}-\ref{eq:lorenzZ}) generate dynamics that can be described as \emph{periodically driven on-off intermittency}.
	In classical on-off intermittency a random or chaotic driving of a system leads to intermittent chaotic behavior, where phases of nearly zero intensity are interrupted by chaotic bursts.
	Typically, the durations of the constant (off) phases follow a power-law distribution.
	Replacing the chaotic driving with a periodic driving leads to alternating nearly constant and chaotic phases with periodic durations.

	In Fig.~\ref{fig:traj}(b) and (d), time series $X(t)$ of the system in Eqs.~\eqref{eq:lorenz} are are shown for two sets of parameter values.
	When the dashed black line appears, $r(t)<1$ is in a regime, where all Jacobian eigenvalues of Eqs.~(\ref{eq:lorenzX}-\ref{eq:lorenzZ}) of the equilibrium $(X,Y,Z)=(0,0,0)$ are negative and thus the trajectory is attracted to this fixed point.
	This equilibrium becomes unstable when the dashed line disappears, where we have $r(t)>1$, and the system reaches a chaotic regime after a transiting through a series of bifurcations \cite{Han2017}.
	Given that the timescale of the parameter variation determined by the period $T$ is sufficiently larger than the correlation time of the system, periodically alternating nearly constant and chaotic phases develop as visible in Fig.~\ref{fig:traj}(b), where the Lyapunov time $\lambda_{\text{max}}^{-1}$ is much smaller than $T$, $\lambda_{\text{max}}^{-1}/T \approx 0.01$ \cite{Note3}.
	In the opposite limit shown in Fig.~\ref{fig:traj}(d), where the Lyapunov time is of the order of $T$, $\lambda_{\text{max}}^{-1}/T \approx 1.18$, the transitions are regular, while the overall dynamics is still chaotic \cite{Note3}.
	Integrating over such time-series, as done by Eq.~\eqref{eq:lorenzx}, leads to nearly constant laminar phases with a chaotic intensity variation and burst-like transitions between them~\footnote{ We note that although multistability is common in many Lorenz-like systems when transitioning from steady states to chaotic states~\cite{Kapitaniak2015,Perks2022}, we do still expect pseudo-laminar chaos to persist in such regimes as long as one varies periodically between a stationary state and a chaotic state}.
	In Fig.~\ref{fig:traj}(a) and (c) the resulting time series $x(t)$ are shown corresponding to Fig.~\ref{fig:traj}(b) and (d), respectively.
	These time series appear to be very similar to so-called \emph{laminar chaos}, which was originally discovered in time-delay systems defined by the delay differential equation (DDE) \cite{muller_laminar_2018}
	\begin{equation}
		\label{eq:dde}
		\dot{x}(t) = -\Theta\,x(t) + \Theta\,f( x\bm{(} R(t) \bm{)}), \quad \text{with } R(t) = t-\tau(t),
	\end{equation}
	where the delay varies periodically, $\tau(t+T) = \tau(t)$.
	In Fig.~\ref{fig:traj}(e) an exemplary laminar chaotic time series is shown, where we have chosen the nonlinearity $f(x) = x + \mu \sin(2\pi\, x)$ and the time-varying delay $\tau(t) = \tau_0 + \frac{A_\tau T}{2\pi} \sin\left(\frac{2\pi}{T}\, t \right)$.
	According to \cite{muller_laminar_2018}, laminar chaos is characterized by nearly constant laminar phases and burst-like transition between them, where the intensity levels of the laminar phases vary chaotically and the durations vary periodically with period $T$, which equals the period of the delay.
	The chaotic dynamics of the laminar phases is governed by the one-dimensional iterated map $x' = f(x)$, which is defined by the nonlinearity of the delayed feedback of Eq.~\eqref{eq:dde}.
	Since all of the time series in Fig.~\ref{fig:traj} are chaotic time series
	\footnote{For Fig.~\ref{fig:traj}(a), (c), and (e) the maximal Lyapunov exponents are given by $\lambda_{\text{max}} \approx 0.24$, $\lambda_{\text{max}} \approx 0.085$, and $\lambda_{\text{max}} \approx 1.26$, respectively.
		The maximal Lyapunov exponents of the parametrically driven Lorenz-like system, Eq.~\eqref{eq:lorenz} were computed using the function \texttt{lyapunovspectrum} from the library \emph{DynamicalSystems.jl} \cite{datseris_dynamicalsystemsjl_2018} and for the time-delay system, Eq.~\eqref{eq:dde}, we adapted the method from \cite{farmer_chaotic_1982}, where the linearized DDE was discretized with step size $\tau_{\text{max}}/M$ with $\tau_{\text{max}}=\max_t \tau(t)$ and $M=2000$.
	}
	 that are characterized by nearly constant phases, which are periodically interrupted by more or less irregular bursts, one may think that they all can be classified as laminar chaos.
	 That would be surprising since laminar chaos is a phenomenon observed in infinite dimensional systems, whereas the system defined by Eq.~\eqref{eq:lorenz} is effectively three-dimensional.
	 In the following we demonstrate that this is indeed not true and that only the time-delay system produces true laminar chaos.
	 We show that there are significant differences between the pseudo-laminar chaotic dynamics shown in Fig.~\ref{fig:traj}(a) and (c) and true laminar chaos shown in (e).
	 We explore in detail the different mechanisms behind pseudo-laminar chaos and laminar chaos and show that pseudo-laminar chaos fails the test for laminar chaos introduced in \cite{hart_laminar_2019,muller-bender_laminar_2020}.
	 Moreover, both types of dynamics react significantly different to additive white noise.

	\section{Test for laminar chaos}
	\label{sec:lamchaostests}
	
	In this section we apply the test for laminar chaos introduced in \cite{hart_laminar_2019,muller-bender_laminar_2020}, and show that it is passed only by the time-series shown in Fig.~\ref{fig:traj}(e), which was generated by the time-delay system, Eq.~\eqref{eq:dde}.
	We start by reviewing some of the basics of the theory of laminar chaos from \cite{muller_laminar_2018,muller-bender_resonant_2019}.
	Equation~\eqref{eq:dde} can be solved using the \emph{method of steps} \cite{bellman_computational_1961,bellman_computational_1965}.
	In this method, the solution $x(t)$ is divided into suitable solution segments $x_k(t)=x(t)$ with $t \in \mathcal{I}_k$, where the boundaries of the \emph{state intervals} $\mathcal{I}_k = (t_{k-1},t_k]$ are connected by the \emph{access map} given by	
	\begin{equation}
		\label{eq:access_map}
		t'=R(t) = t-\tau(t).
	\end{equation}
	Assuming that $R(t)$ is monotonically increasing, i.e., $\dot{\tau}(t)<1$ for almost all $t$, which can be motivated by physical and mathematical arguments \cite{verriest_inconsistencies_2011,verriest_state_2012}, the state intervals cover the whole time-domain and the $k$th solution segment $x_k(t)$ can be interpreted as the memory of the delay system at time $t_k$.
	Given an initial segment $x_0(t)$, the dynamics of the solution segments is governed by an iteration of a nonlinear operator, where subsequent segments are connected by
	\begin{equation}
		\label{eq:soluop}
		x_{k+1}(t) = x_{k}(t_{k}) e^{-\Theta(t-t_{k})} + \int\limits_{t_{k}}^{t} \! dt' \, \Theta e^{-\Theta(t-t')} f(x_k\bm{(}R(t')\bm{)}).
	\end{equation}
	Since the integral kernel $\Theta e^{-\Theta(t-t')}$ can be interpreted as an approximation of the delta distribution with width $\Theta^{-1}$ \cite{ikeda_high-dimensional_1987}, one has, in the limit $\Theta\to\infty$,
	\begin{equation}
		\label{eq:limit_map}
		x_{k+1}(t) = f(x_k\bm{(}R(t)\bm{)}),
	\end{equation}
	which can also be derived directly from Eq.~\eqref{eq:dde} by dividing both sides by $\Theta$, taking the limit $\Theta\to\infty$ and replacing $x(t)$ and $x(R(t))$ with $x_{k+1}(t)$ and $x_k(R(t))$, respectively.
	Moreover, Eq.~\eqref{eq:soluop} can be interpreted as application of the so-called limit map, Eq.~\eqref{eq:limit_map}, and a subsequent smoothing with a kernel of width $\Theta^{-1}$.
	Since the influence of smoothing can be neglected inside the nearly constant laminar phases, given that the width of the kernel is much smaller than the duration of the laminar phases, in the laminar chaotic regime, the dynamics of Eq.~\eqref{eq:dde} can be well approximated by Eq.~\eqref{eq:limit_map}.
	The limit map, Eq.~\eqref{eq:limit_map} can be understood as the pointwise iteration of the graph $(t,x_k(t))$, with $t \in \mathcal{I}_{k}=(t_{k-1},t_{k}]$, under the two-dimensional map
	\begin{subequations}
		\label{eq:2d_map}
		\begin{align}
			y_k &= R^{-1}(y_{k-1}) \label{eq:2d_map_r}\\
			z_k &= f(z_{k-1}) \label{eq:2d_map_f},
		\end{align}
	\end{subequations}
	which consists of two independent one-dimensional maps:
	Equation~\eqref{eq:2d_map_f} describes the dynamics of the function values from solution segment to solution segment and thus it governs the dynamics of the intensity levels of the laminar phases.
	Equation~\eqref{eq:2d_map_r} acts on the time-domain and describes the frequency modulation of the solution segments caused by the temporal variation of the delay $\tau(t)$.
	Since the delay system described by Eq.~\eqref{eq:dde} is a feedback loop, this kind of frequency modulation occurs with each roundtrip inside the feedback loop, where the memory of the system after the $k$th roundtrip is represented by the solution segment $x_k(t)$.
	If there is a resonance between the average roundtrip time inside the feedback loop and the period of the delay variation, periodically alternating low- and high-frequency phases develop, where the low-frequency phases degenerate to nearly constant phases for laminar chaos.
	Mathematically the roundtrip of the signal inside a feedback loop with time-varying delay is described by Eq.~\eqref{eq:2d_map_r} and its inverse, the access map given by Eq.~\eqref{eq:access_map}.
	For periodically time-varying delays they are lifts of circle maps (cf. \cite{arnold_small_1961,*arnold_small_1961_erratum,katok_introduction_1997}) and a resonance between the rountrip inside the feedback loop and the delay variation corresponds to stable mode-locking dynamics of these circle maps, i.e., in this case the Lyapunov exponent $\lambda[R]$ of the access map is negative.
	Analyzing the stability of a piecewise constant solution of Eq.~\eqref{eq:limit_map} leads to the conclusion that laminar chaos can be observed if the condition
	\begin{equation}
		\label{eq:lamchaos_crit}
		\lambda[f] + \lambda[R] < 0
	\end{equation}
	is fulfilled, where $\lambda[f]$ is the Lyapunov exponent of the map, Eq.~\eqref{eq:2d_map_f}, defined by the nonlinearity of the delayed feedback.
	A generalization of the theory of laminar chaos to systems with quasiperiodic delay can be found in \cite{muller-bender_laminar_2022}.
	Laminar chaos was also found in system with a more complex structure than Eq.~\eqref{eq:dde}:
	The synchronization of laminar chaotic systems was considered in \cite{khatun_synchronization_2022} and, as shown in \cite{kulminskiy_laminar_2022}, laminar chaos can be induced into a constant delay system if it is coupled to a time-varying delay system.
	
	\begin{figure}
		\includegraphics[width=0.48\textwidth]{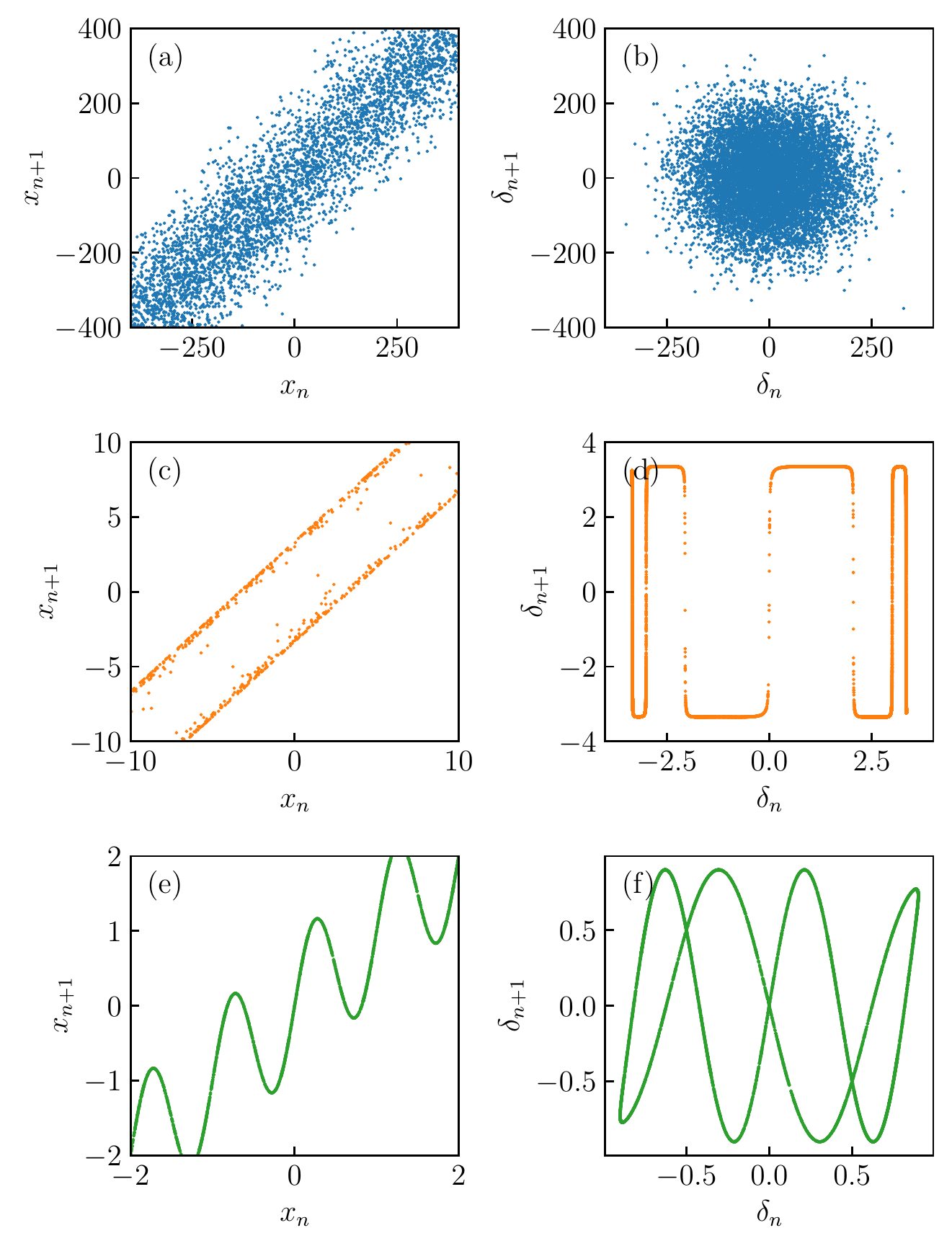}
		\caption{Return maps of the levels $x_n$ of the (pseudo) laminar phases and of the increments $\delta_n = x_{n+1} - x_n$.
			The parameters were chosen as in Fig.~\ref{fig:traj}, where (a) and (b) correspond to Fig.~\ref{fig:traj}(a), (c) and (d) correspond to Fig.~\ref{fig:traj}(c), (e) and (f) correspond to Fig.~\ref{fig:traj}(e) as indicated by the colors.
			The levels are estimated by setting $x_n=x(n\,T)$, where $T$ is the period of $r(t)$ or of the time-varying delay $\tau(t)$ for pseudo-laminar and laminar chaos, respectively.
			For the laminar chaos considered in (e) and (f), the time-instants $t^*_n=n\,T$ are the members of the drifting attractive orbits of the access map, Eq.~\eqref{eq:access_map} (see text).
		}
		\label{fig:retmaps}
	\end{figure}

	\begin{figure}
		\includegraphics[width=0.48\textwidth]{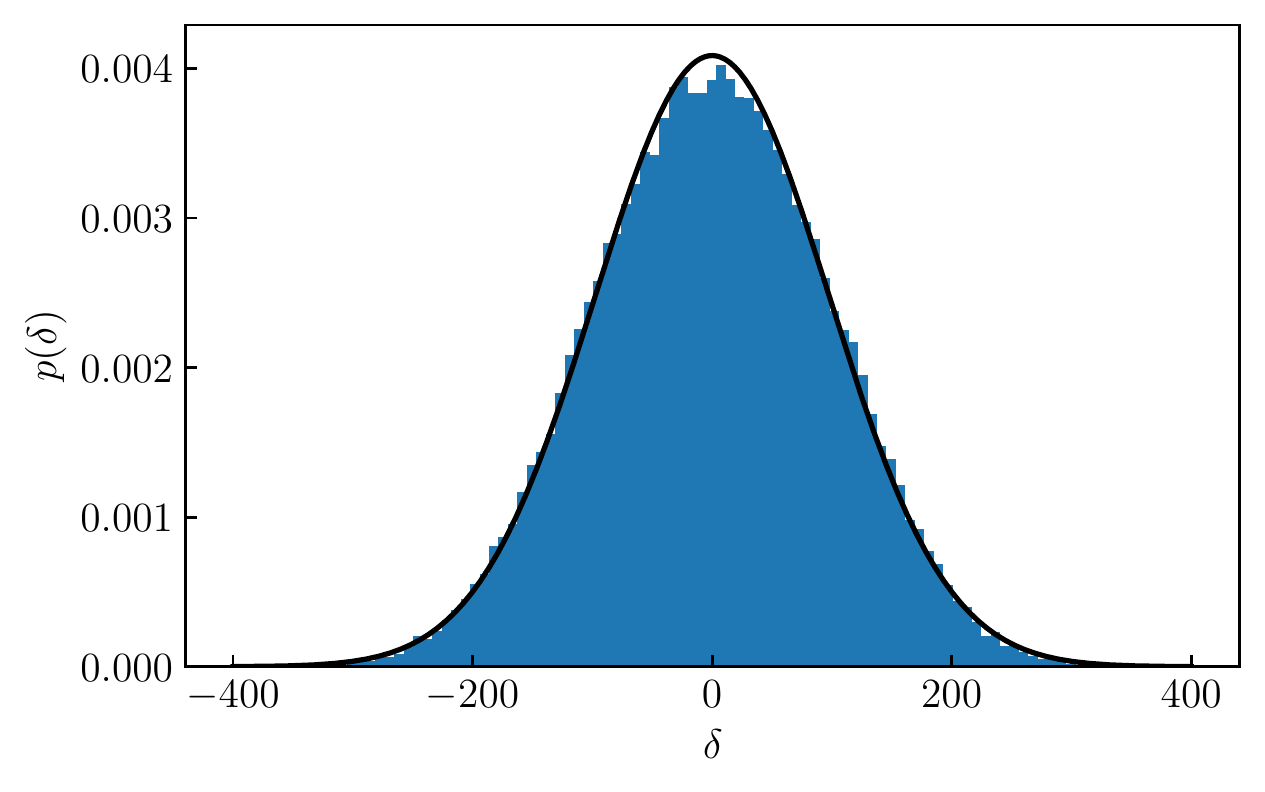}
		\caption{Histogram of the increments $\delta_n = x_{n+1} - x_n$ shown in Fig.~\eqref{fig:retmaps}(b) and Gaussian approximation (solid line) with zero mean, where the standard deviation was computed from the data.  
		}
		\label{fig:inchist}
	\end{figure}
	
	According to \cite{muller-bender_laminar_2020}, the test for laminar chaotic dynamics consists of the detection of two features, which are robust against noise so that they can be detected even for large noise strengths, where nearly constant phases are hard to recognize.
	Only if both features are present, the dynamics can be classified as laminar chaos.
	The first feature consists of laminar phases with periodic durations, which appear as periodically repeating low amplitude phases of the derivative $\dot{x}(t)$ of the signal $x(t)$.
	These phases are present for all considered time series as illustrated in Fig.~\ref{fig:traj}(b), (d), and (f).
	The intensity levels $x_n$ of the laminar phases are measured, which are numbered by the index $n$ in the same order as they appear in the time series.
	A natural approach for measuring the intensity levels is setting $x_n=x(t^*_n)$, where the $t^*_n$ are members of the drifting attractive orbits of the access map, Eq.~\eqref{eq:access_map}.
	The time instants $t^*_n$ are crucial for the long time dynamics of the intensity levels since, for large $k$, we have inside a laminar phase $x_k(t) = f^k(x_0\bm{(}R^k(t)\bm{)}) \approx f^k(x_0(t^*_n))$ according to Eq.~\eqref{eq:limit_map}, where $t^*_n$ is an attractive point inside the initial state interval $\mathcal{I}_0$.
	When one considers systems with noise, fluctuations of $x(t)$ are minimal at the $t^*_n$ (cf. \cite{muller-bender_laminar_2020}), which is convenient for detecting the second robust feature of laminar chaos given by the fact that the dynamics of the intensity levels $x_n$ is governed by the one-dimensional map defined by Eq.~\eqref{eq:2d_map_f}.
	In particular we have $x_{n+p}=f(x_n)$, where $p$ is the number of laminar phases per state interval $\mathcal{I}_k$, which follows directly from the required monotonicity of the access map, Eq.~\eqref{eq:access_map}:
	Since the access map is monotonically increasing, it is clear that the first laminar phase in $\mathcal{I}_k$ is mapped by Eq.~\eqref{eq:soluop} to the first laminar phase in $\mathcal{I}_{k+1}$, the second to the second, and so forth.
	This feature can be detected by trying to reconstruct the nonlinearity $f$ from the intensity levels $x_n$, where we plot $x_{n+p'}$ over $x_n$.
	If we find a positive integer $p'$ such that the points $(x_n,x_{n+p'})$ resemble a graph of a function, the dynamics is classified as laminar chaos, where the function is an iteration $f^{l}(x)$ of the right hand side of Eq.~\eqref{eq:2d_map_f}.
	For the smallest $p'=p$, the nonlinearity $f$ of the delayed feedback is reconstructed.
	In Fig.~\ref{fig:retmaps}(a), (c), and (e) such return maps with $p'=1$ are shown for time series of Eq.~\eqref{eq:lorenz} and Eq.~\eqref{eq:dde}, where the parameters are chosen as in Fig.~\ref{fig:traj} as indicated by the coloring.
	While for the time series generated by Eq.~\eqref{eq:lorenz} no $p'$ was found with the property that the points $(x_n,x_{n+p'})$ resemble a function, for the delay system defined by Eq.~\eqref{eq:dde} the reconstructed nonlinearity $f(x)$ is clearly visible in Fig.~\ref{fig:retmaps}(e).
	It follows that the exemplary time-series of Eq.~\eqref{eq:dde} can be classified as laminar chaos, whereas the two exemplary time-series of Eq.~\eqref{eq:lorenz} are pseudo-laminar chaos since they fail the test for laminar chaos.
	Similar results are obtained if the sinusoidal variation of the parameter $r$ in Eqs.~\eqref{eq:lorenz} is replaced by a periodic, piecewise constant function that switches between $+A_r$ and $-A_r$ as shown in Appendix~\ref{sec:app_switched}.
	These qualitative observations can be verified by quantifying nonlinear correlations of the return maps of the levels $x_n$ as done in Appendix~\ref{sec:app_corr}, where the results of two methods are compared.
	For all considered parameter values, both methods lead to a clear distinction between pseudo-laminar and true laminar chaos.
    To summarize, while all considered time series show nearly constant laminar phases with chaotically varying levels, only the time series generated by the delay system, Eq.~\eqref{eq:dde}, satisfy the central property of true laminar chaos: The dynamics of the levels is governed by a one-dimensional iterated map.
    This property is not satisfied by the time series generated by Eqs.~\eqref{eq:lorenz} and therefore we call the dynamics pseudo-laminar chaos.
	
	If one considers the increments $\delta_n = x_{n+p'}-x_n$ instead of the levels $x_n$, which seems to be a reasonable approach since the $\delta_n$ are bounded whereas the $x_n$ are unbounded for chaotic diffusion, the picture is not that clear.
	While the return map of the increments shown in Fig.~\ref{fig:retmaps}(b) appears clearly different from the return map obtained for laminar chaos, Fig.~\ref{fig:retmaps}(f), in Fig.~\ref{fig:retmaps}(d) line-like structures appear, which are somewhat similar to the structures obtained for laminar chaos in Fig.~\ref{fig:retmaps}(f).
	The numerical analysis of the nonlinear correlations of these return maps in Appendix~\ref{sec:app_corr} confirms these qualitative observations:
	While the quantitative analysis for Fig.~\ref{fig:retmaps}(b) indicates the absence of nonlinear correlations, strong correlations are found in Fig.~\ref{fig:retmaps}(d) similar to Fig.~\ref{fig:retmaps}(f), although only the latter one corresponds to true laminar chaos.
	As a consequence, one should be aware that only the return map of the levels of the laminar phases should be used for the detection of laminar chaos and not the return map of the increments.
	However, as we demonstrate in Sec.~\ref{sec:noise}, the line-like structures in Fig.~\ref{fig:retmaps}(d) disappear in the presence of noise, whereas the features of true laminar chaos are robust against noise as demonstrated in \cite{hart_laminar_2019,muller-bender_laminar_2020}.
	
	The differences of the structures in the return maps observed for laminar and pseudo-laminar chaos are direct consequences of the differences between the mechanisms behind these dynamics.
	For laminar chaos the one-dimensional map that generates the levels $x_n$ is induced by the nonlinear delayed feedback:
	If the condition for laminar chaos, Eq.~\eqref{eq:lamchaos_crit}, is fulfilled and $\Theta$ is large enough, the dynamics in the laminar phases can be well described by the two-dimensional map, Eq.~\eqref{eq:2d_map}.
	At the transition between these phases, this approximation breaks down, since there the smoothing in Eq.~\eqref{eq:soluop} becomes relevant.
	In this sense, the dynamics of the levels of the laminar phases is independent of the dynamics of the transitions in between.
	In strong contrast, for the pseudo-laminar chaos introduced here, the level dynamics are dictated by the transitions since pseudo-laminar chaos is the integral of chaotic bursts according to Eq.~\eqref{eq:lorenzx}, and thus each increment $\delta_n = x_{n+1}-x_n$ is given by the integral of a chaotic burst between two subsequent laminar phases.
	Assuming that the duration of the chaotic bursts are large compared to the correlation length of the chaotic oscillations in the chaotic burst phase, the increments $\delta_n$ can be viewed as a sum of a large number of nearly independent pseudo-random numbers.
	As a result in this limit, the increments $\delta_n$ are nearly uncorrelated as visible in Fig.~\ref{fig:retmaps}(b) and, according to the central limit theorem, are nearly Gaussian distributed as illustrated in Fig.~\ref{fig:inchist}, where a histogram of $\delta_n$ for the parameters used in Fig.~\ref{fig:traj}(a), (b), and Fig.~\ref{fig:retmaps}(a), (b) is shown together with a Gaussian approximation.
	Line-like structures in the increments, which imply strong (nonlinear) correlations and a non-Gaussian distribution in general, as shown in Fig.~\ref{fig:retmaps}(d) can be obtained if one tunes the parameters such that the duration of the bursts is close to the correlation time of the chaotic oscillations so that each burst consists of one short oscillation as visible in Fig.~\ref{fig:traj}(c).
	As one may expect, this kind of behavior is not robust, i.e., it is very sensitive to noise.
	We explore this in the next section.

	\section{Sensitivity to noise}
	\label{sec:noise}
	
	\begin{figure}[h]
		\includegraphics[width=0.48\textwidth]{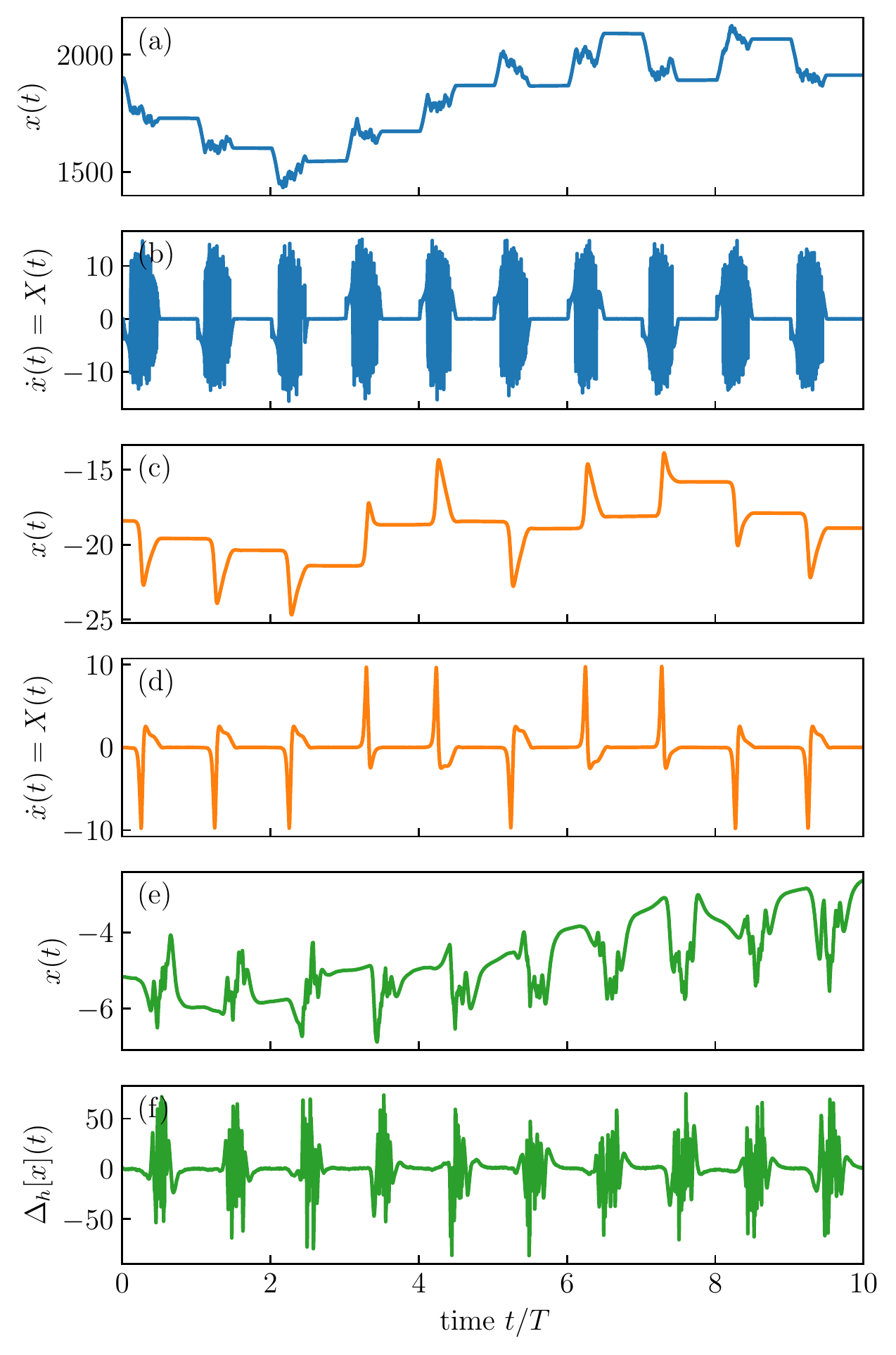}
		\caption{Effect of dynamical noise on time series of pseudo-laminar chaos and laminar chaos.
			In (a) and (c) pseudo-laminar chaos and in (e) true laminar chaos is shown, where additive Gaussian white noise with a small standard deviation $\epsilon=0.01$ was added to the right hand side of Eq.~\eqref{eq:lorenzX} for pseudo-laminar chaos and to Eq.~\eqref{eq:dde} for laminar chaos.
			The corresponding derivatives of the time series are plotted below in Figs.~(b),(d), and (f), where, for laminar chaos, the approximate derivative $\Delta_h(x)(t)=h^{-1} [x(t+h)-x(t)]$ with $h\approx 0.0023$ is shown.
			The parameters were chosen as in Fig.~\ref{fig:traj}.
		}
		\label{fig:trajnoise}
	\end{figure}

	\begin{figure}[h]
		\includegraphics[width=0.48\textwidth]{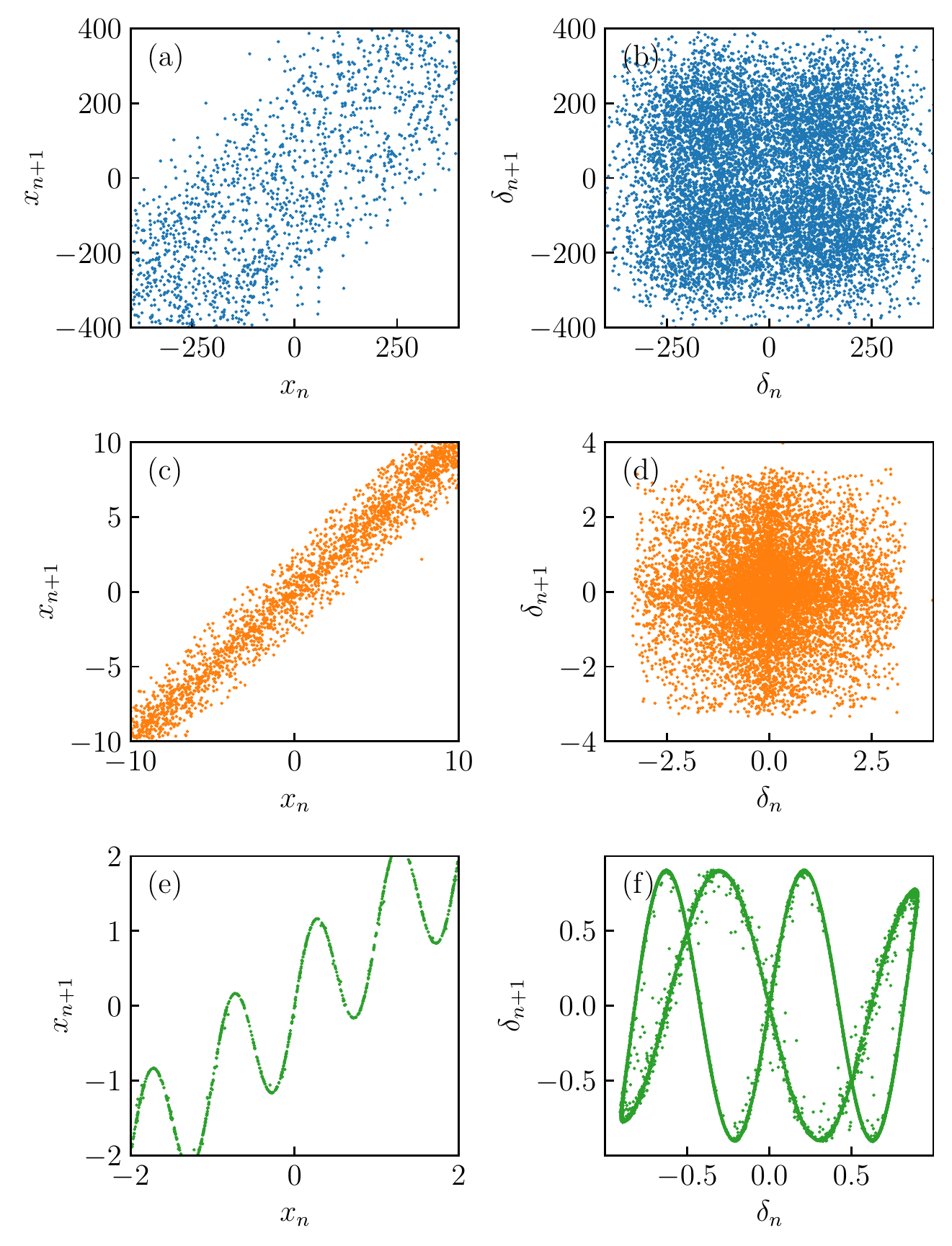}
		\caption{Return maps of the levels $x_n$ of the (pseudo) laminar phases and of the increments $\delta_n = x_{n+1} - x_n$ in systems with noise.
			While the features of laminar chaos shown in (e) and (f) persist, the line-like structures obtained for pseudo-laminar chaos, which are shown in Fig.~\ref{fig:retmaps}(c) and (d), are destroyed by the noise as illustrated in (c) and (d).
			The parameters were chosen as in Fig.~\ref{fig:traj}, where (a) and (b) correspond to Fig.~\ref{fig:traj}(a), (c) and (d) correspond to Fig.~\ref{fig:traj}(c), (e) and (f) correspond to Fig.~\ref{fig:traj}(e) as indicated by the colors.
			Additive Gaussian white noise with a small standard deviation $\epsilon=0.01$ was added to the right hand side of Eq.~\eqref{eq:lorenzX} for pseudo-laminar chaos and to Eq.~\eqref{eq:dde} for laminar chaos.
			The levels are estimated by setting $x_n=x(n\,T)$, where $T$ is the period of $r(t)$ or of the time-varying delay $\tau(t)$ for pseudo-laminar and laminar chaos, respectively.
			For the laminar chaos considered in (e) and (f), the time-instants $t^*_n=n\,T$ are the members of the drifting attractive orbits of the access map, Eq.~\eqref{eq:access_map}, where the fluctuations due to the noise are minimal (see Sec.~\ref{sec:lamchaostests}).
		}
		\label{fig:retmapsnoise}
	\end{figure}

	In this section we analyze the influence of additive white noise on pseudo-laminar chaos and laminar chaos.
	We demonstrate that pseudo-laminar chaos reacts fundamentally different to noise than laminar chaos.
	While it is known that the robust features of laminar chaos considered in Sec~\ref{sec:lamchaostests} can survive for larger noise strengths \cite{muller-bender_laminar_2020}, it turns out for pseudo-laminar chaos that even small noise strengths can change properties such as line-like structures in the return map of the increments drastically.
	In order to do this we add white noise to Eq.~\eqref{eq:lorenz} and Eq.~\eqref{eq:dde}.
	We then perform the same analysis as in Sec.~\ref{sec:lamchaostests} and compare the results.
	For simplicity we add noise only to the right hand side of Eq.~\eqref{eq:lorenzX}.
	So we consider the stochastic differential equation
	\begin{subequations}
		\begin{align}
			\dot{x}(t) &= X(t) \label{eq:lorenznoisex}\\
			\dot{X}(t) &= \sigma\, [Y(t) - X(t)] + \epsilon\, \xi(t) \label{eq:lorenznoiseX}\\
			\dot{Y}(t) &= X(t)\, [r - Z(t)] - Y(t) \label{eq:lorenznoiseY}\\
			\dot{Z}(t) &= X(t)\,Y(t) - b\,Z(t) \label{eq:lorenznoiseZ},
		\end{align}
	\label{eq:lorenznoise}
	\end{subequations}
	for pseudo-laminar chaos, where qualitatively equivalent results are obtained if noise is added to another or to more than one of the Eqs.~(\ref{eq:lorenzX}-\ref{eq:lorenzZ}).
    We also expect equivalent results if colored noise with a correlation length much smaller than the period of the variation of the parameter $r=r(t)$ is considered.
	Adding noise to Eq.~\eqref{eq:lorenzx} is less interesting since this noise does not feed back to the dynamical system and it is basically the same as adding Brownian motion to $x(t)$.
	For laminar chaos we consider the stochastic DDE
	\begin{equation}
		\label{eq:ddenoise}
		\dot{x}(t) = -\Theta\,x(t) + \Theta\, f( x\bm{(} R(t) \bm{)}) + \epsilon\, \xi(t),
	\end{equation}
	where $\xi(t)$ is Gaussian white noise with mean $\langle \xi(t) \rangle = 0$ and covariance function $\langle \xi(t) \xi(t') \rangle = \delta(t-t')$.
	
	In Fig.~\ref{fig:trajnoise} time series $x(t)$ and their derivatives $\dot{x}(t)$ are shown for these systems, where we have set $\epsilon = 0.01$ and used the same parameters as in Fig.~\ref{fig:traj}.
	The corresponding return maps of the levels $x_n$ of the laminar phases and of their increments $\delta_n$ are shown in Fig.~\ref{fig:retmapsnoise}.
	We find that in the presence of noise laminar phases are visible for both, pseudo-laminar chaos and laminar chaos.
	At first sight, the laminar phases of true laminar chaos, Fig.~\ref{fig:trajnoise}(e) appear more perturbed than the laminar phases of pseudo-laminar chaos, Fig.~\ref{fig:trajnoise}(a) and (c). 
	However, despite the deviations from the nearly constant behavior of the laminar phases in the noiseless system, Fig.~\ref{fig:traj}(e), the nonlinearity of the feedback $f$ can be nicely reconstructed as shown in Fig.~\ref{fig:retmapsnoise}(e), which is also reflected by the return map of the increments shown in Fig.~\ref{fig:retmapsnoise}(f) and confirms the robustness of this feature of laminar chaos against noise.
	In strong contrast, the line-like behavior of the return map of the increments observed for pseudo-laminar chaos in Fig.~\ref{fig:retmaps}(d) is completely destroyed.
	Moreover, the structures observed in the return maps become similar to the ones observed for the limiting case, Fig.~\ref{fig:retmapsnoise}(a) and (b), where the transitions between the laminar phases are given by the integration over many chaotic fluctuations.
	This indicates that such a behavior is typical for pseudo-laminar chaos.
	In other words, the line-like behavior shown by the noiseless system is not robust and thus it is unlikely to observe it in experiments unless the noise strength is sufficiently small.
	These observations are confirmed by the analysis in Appendix~\ref{sec:app_corr}, where nonlinear correlations of the return maps are quantified for the considered set of parameter values.
	While correlations detected in the noiseless system persist in the presence of noise for true laminar chaos, for pseudo-laminar chaos, the correlations visible in Fig.~\ref{fig:retmaps}(d) vanish in the presence of noise.
	
	The drastic differences between pseudo-laminar chaos and laminar chaos with respect to the reaction to noise reflect the drastic differences of the generating mechanisms of the related laminar phases.
	During the laminar phases of pseudo-laminar chaos, noise drives the subsystem Eqs.~\eqref{eq:lorenzX}-\eqref{eq:lorenzZ} away from the equilibrium $(X,Y,Z)=(0,0,0)$, which is stable during these phases.
	This leads to larger side peaks of $\dot{x}(t)=X(t)$ in Fig.~\ref{fig:trajnoise}(d) in comparison to the small side peaks in Fig.~\ref{fig:traj}(d).
	Since the increments of the intensity levels are given by the integral over $X(t)$ these deviations are even more amplified, so that potential low-dimensional structures disappear from the return map as reflected by the differences between Fig.~\ref{fig:retmaps}(c), (d) and Fig.~\ref{fig:retmapsnoise}(c), (d).
	In contrast, a delay system given by Eq.~\eqref{eq:dde}, which can generate true laminar chaos, reacts in a completely different way to additive noise, which is a consequence of the infinite dimensional phase space of the delay system, whereas the system defined by Eq.~\eqref{eq:lorenz} is finite dimensional.
	Using the definitions of the solution segments from Sec.~\ref{sec:lamchaostests}, the solution operator of the stochastic system given by Eq.~\eqref{eq:ddenoise} reads
	\begin{align}
		\label{eq:soluopnoise}
		x_{k+1}(t) = x_{k}(t_{k}) & e^{-\Theta(t-t_{k})} \\
		&+ \int\limits_{t_{k}}^{t} \! dt' \, \Theta e^{-\Theta(t-t')} f(x_k\bm{(}R(t')\bm{)}) \nonumber \\
		&+ \epsilon \Xi_k(t) \nonumber,
	\end{align}
	which differs from the solution operator, Eq.~\eqref{eq:soluop}, of the noiseless system by the random function segment $\Xi_k(t)$ defined by
	\begin{equation}
		\Xi_k(t) = \int\limits_{t_{k}}^{t} \! dt' \, e^{-\Theta(t-t')} \, \xi(t').
	\end{equation}
	So $\Xi_k(t)$ is generated by an Ornstein-Uhlenbeck process with initial value $\Xi_k(t_k)=0$.
	For large $\Theta$, we can neglect the influence of the smoothing operator in Eq.~\eqref{eq:soluopnoise} leading to the approximation
	\begin{equation}
		x_{k+1}(t) = f(x_k\bm{(}R(t)\bm{)}) + \epsilon \Xi_k(t),
	\end{equation}
	which can be viewed as a stochastic version of the limit map, Eq.~\eqref{eq:limit_map}, and thus it can be interpreted as the iteration of the graph $(t,x_k(t))$ under the two-dimensional map
	\begin{subequations}
		\label{eq:2d_mapnoise}
		\begin{align}
			y_k &= R^{-1}(y_{k-1}) \label{eq:2d_mapnoise_r}\\
			z_k &= f(z_{k-1}) + \epsilon \Xi_k(y_k)  \label{eq:2d_mapnoise_f},
		\end{align}
	\end{subequations}
	where the dynamics of the function values of the segments, and thus the dynamics of the levels of the laminar phases, is governed by Eq.~\eqref{eq:2d_mapnoise_f}.
	So from this first order approximation it follows that additive white noise in the delay system leads to additive noise in the map, Eq.~\eqref{eq:2d_mapnoise_f}, generating the levels of the laminar phases, where the mechanism behind and thus the features of laminar chaos remain intact.
	For larger noise strengths, where the smoothing by the kernel $\Theta e^{-\Theta(t-t')}$ cannot be neglected, it was demonstrated experimentally and numerically in \cite{hart_laminar_2019,muller-bender_laminar_2020} that the nonlinearity $f$ of the delayed feedback can be reconstructed even if the laminar phases are hard to recognize, which verifies the robustness of laminar chaos.

	\section{Discussion}
	\label{sec:discussion}
	
	   It would be interesting to investigate whether pseudo-laminar chaos can be realized in experiments for example in the hydrodynamical system of single-frequency driven walkers and two-frequency driven superwalkers. One way this could be achieved is by periodically driving the droplets between a stationary (non-walking) state similar to a laminar phase and a chaotic walking state similar to a chaotic burst. The stop-and-go motion of superwalkers~\citep{superwalker,superwalkernumerical} arising from slight detuning of two driving frequencies, provides a convenient way to periodically drive the system between a stationary state and a walking state. However, only steady walking states have been observed in experiments with a superwalker in free space. For a walker in free space, in addition to steady walking states, oscillatory walking states have also been reported in experiments~\citep{Bacot2019}. However, chaotic walking regimes that are predicted to arise in theoretical models~\citep{Hubert2019,Durey2020lorenz,Durey_2D_2021,ValaniUnsteady,Valanilorenz2022}, have not been realized in experiments yet. Nevertheless, chaotic walking states have been observed for a walker in confining potentials~\citep{Chaosmemoryperrard,Perrard2014a,Tudor2018,PhysRevE.88.011001} and rotating frames~\citep{Harris2013}, and similar chaotic states might be expected for superwalkers in confining potentials. Hence, if one investigates stop-and-go motion of superwalkers in confining potentials, then one may be able to oscillate between a stationary and chaotic walking states, and pseudo-laminar chaos may be realized in such an experimental setup.
	   
	   Although our investigations of pseudo-laminar chaos were inspired by models of walking droplets with a periodic driving, our results are obviously more general: any system showing periodic on-off intermittency would show signatures of laminar chaos in the time-integrated intermittent variable, but a proper analysis, as done in this paper, would reveal that one has observed actually pseudo-laminar chaos instead of true laminar chaos. There are many reports on experiments \cite{hammer_experimental_1994,john_on-off_1999,cabrera_on-off_2002} and numerical simulations \cite{yang_on-off_1994,yang_on-off_1996,cabrera_on-off_2002} on driven on-off intermittency, usually with a random driving. Obviously the latter could be changed to a periodic driving, which then would produce examples of pseudo-laminar chaos in the integrated intermittent variable, as reported here for the walking droplet dynamics, which led to an integrated Lorenz-like dynamical system. Potential candidates of systems where pseudo-laminar chaos can be induced are given by systems that show chaotic deterministic diffusion \cite{klages_deterministic_1996}. In such systems, the velocity shows bounded chaotic dynamics and its time-integral, i.e., the state variable shows diffusion. Let us consider such a system with an experimentally accessable parameter, which can be tuned such that the velocity shows chaos for one parameter value and converges to a stable equilibrium equal to zero for another value. If one periodically switches the parameter value between these regimes, the velocity shows periodically driven on-off intermittency given that the period of the switching is large compared to the correlation time of the system. As a result, the state variable shows pseudo-laminar chaos. This concept can be easily applied to the periodically forced nonlinear pendulum considered in \cite{dhumieres_chaotic_1982}, which is also a model for phase-locked loops and Josephson junctions \cite{huberman_noise_1980,schell_diffusive_1982}. If the external periodic driving is periodically switched on and off, where the pendulum rotates chaotically in the on phase and relaxes to the equilibrium in the off phase, pseudo laminar chaos is observed in the angular position. A subclass of chaotic deterministic diffusion is the so-called deterministic Brownian motion \cite{trefan_deterministic_1992,mackey_deterministic_2006}, which is also found in time-delay systems where the velocity shows turbulent chaos \cite{lei_deterministic_2011}. Therefore, pseudo-laminar chaos can even be induced in time-delay systems by periodically switching the dynamics of the velocity between turbulent chaos and relaxation to a stable equilibrium equal to zero. This highlights the relevance of our test for distinctive features of pseudo-laminar and true laminar chaos provided in Sec.~\ref{sec:lamchaostests}, especially in experimental situations where the mechanism generating the measured time series is unknown.
    
On the other hand, on-off intermittency in the narrow sense as formulated originally in \cite{platt_on-off_1993,heagy_characterization_1994}, is connected to the appearance of laminar phases, whose durations are not roughly constant as with periodic driving, but power law distributed. In this case, the time-integrated intermittent variable, would consist of a sequence of randomly varying plateaus, with durations governed by the same power-law distribution. In such cases one would observe a generalized version of pseudo-laminar chaos. A similar dynamical behavior that is also based on on-off intermittency is the so-called \emph{multistate on-off intermittency} \cite{lai_intermingled_1995,zhan_intermingled_2000} and the dynamics shown by the phase difference of coupled periodically driven pendula close to the onset of synchronization \cite{yang_transition_2001}. If one allows for other statistical variations of the durations, such generalized pseudo-laminar chaos could also be observed for walking droplets. For example, in the chaotic walking regimes realized in theoretical models of walking droplets in two-dimensions, a run-and-tumble-like trajectory is observed~\citep{Hubert2019,Durey_2D_2021}. For such trajectories, if one plots the time series of the particle's orientation then it would be a constant during the run phase (laminar phase) and vary chaotically during the tumble phase. Hence, the durations of the run phases determine the statistics of laminar phases, while the random variations of the angle in the tumble phases determines the increments between the angles of subsequent laminar phases. The time dependence of the angle is naturally monitored as a time-continuous real variable without discontinuities after full revolutions, thus leading to a phase diffusion process, which may be normal or anomalous depending on the duration statistics of the walking phases. Moreover, this more general idea of pseudo-laminar chaos could also be transferred to stochastic processes such as Lévy flights~\cite{1995} where a trajectory may be partitioned into a ``laminar" phase of a long step and a ``chaotic" burst of small steps. This could be relevant for the description of human stick balancing, where a combination of on-off intermittency and Lévy flights was found \cite{cabrera_human_2004}.

So far all the discussed pseudo-laminar chaotic processes, were actually diffusion processes. They arise naturally because integrating over chaotic bursts leads to independent random increments. It appears to be difficult to generate bounded pseudo-laminar chaotic processes because this would need systematic correlations between subsequent chaotic bursts, which is not to be expected for on-off intermittency. In contrast, for true laminar chaos a bounded signal appears naturally, as in systems studied in~\citep{muller_laminar_2018}, but also true laminar chaotic diffusion with an unbounded signal variation arises naturally \cite{albers_chaotic_2022}. 
As an outlook we may wonder, whether, as a counterpart to generalized pseudo-laminar chaos as discussed above, true laminar chaos from delay systems can also show randomly varying durations of the laminar phase, at variance with the periodic variation obtained in ~\citep{muller_laminar_2018}. The tentative answer is affirmative: a random variation of these durations is expected to occur for delay systems with randomly varying delay. As a first step, it was confirmed that the random counterparts of circle maps, which arise from the delayed argument, typically show mode-locking \cite{muller-bender_suppression_2022}, which is a condition to be fulfilled for the existence of laminar chaos in delay systems with random delay. Preliminary investigations of such delay systems show that indeed laminar chaos can exist, but the statistics of the durations of the laminar phases has still to be explored. Another hint for the existence of distributed durations comes from studies of delay systems with a delay intermediate between periodic and random, namely systems with a quasiperiodic delay variation. There we found \cite{muller-bender_laminar_2022}, that indeed laminar chaos is possible and that it becomes more frequent if the quasiperiodic variation approaches a random variation. The differences between generalized pseudo-laminar chaos and laminar chaos from delay systems with random delay are expected to be similar to those in periodic systems as found in this work: again the infinitely many negative Lyapunov exponents of the delay systems have a strongly stabilizing effect against noise, which is missing in systems with only a few degrees of freedom.

Lastly, although we considered a simple model that reduces to a finite-dimensional Lorenz-like ODEs (Eqs.~\eqref{eq:lorenz}) for a walking-droplet inspired 1D wave-particle entity with a sinusoidal wave field, a more refined model, as originally proposed by \citet{Oza2013}, which more accurately captures the experimentally observed wave field in the form of a Bessel function results in an integro-differential equation of motion for the wave-particle entity. An integro-differential equation of motion also arises if one considers the wave-particle entity in two-dimensions. The integro-differential equation is a result of the path-memory in the hydrodynamic system since the motion of the droplet is not only influenced by its most recently generated wave but also by the waves generated in the distant past. Hence, one may wonder whether true laminar chaos might be observed in such walking-droplet inspired path-memory induced integro-differential equations or more generally in delay systems with state-dependent delay where also infinite degrees of freedom are present. 
	
	\section{Summary}
	
	We investigated diffusive time series of a finite-dimensional, chaotic, parametrically driven Lorenz-like dynamical system, which can be motivated, for instance, by a hydrodynamical wave-particle system.
	While these time series share their appearance with that of recently discovered laminar chaotic diffusion \cite{albers_chaotic_2022}, they fail the test for laminar chaos proposed in \cite{muller-bender_laminar_2020}.
	It turned out that such pseudo-laminar chaos is generated by a mechanism completely different from that of laminar chaos.
	Laminar chaos is found for systems with nonlinear feedback-loops with a time-varying delay, which are infinite dimensional systems described by singularly perturbed delay differential equations.
	The interplay between the map defined by the nonlinearity of the feedback and the map defined by the time-varying delay leads to the development of nearly constant laminar phases with periodic durations, where the intensity levels of the laminar phases vary chaotically from phase to phase and follow the dynamics of the chaotic one-dimensional map defined by the nonlinearity of the feedback.
	In strong contrast, pseudo-laminar chaos is generated by integrated periodically driven on-off intermittency.
	While the latter is also characterized by nearly constant laminar phases with chaotically varying intensity and periodic durations, the intensity levels do not follow a one-dimensional iterated map.
	Moreover, if the correlation time of the intermittent signal is much smaller than the durations of the chaotic bursts between the laminar phases, the intensity levels rather follow a random walk, since their increments, which are the integral over the bursts between subsequent laminar phases, are nearly uncorrelated.
	As a consequence of the different mechanisms and the drastically different dimensionality of the involved systems, we found fundamental differences in the reaction to adding white noise in the equations of motion.
	As shown in \cite{hart_laminar_2019,muller-bender_laminar_2020}, true laminar chaos is robust to noise in the sense that its characteristic features such as the nonlinear correlations between the laminar phases can be detected even for larger noise strength, where the laminar phases are hardly visible.
	In contrast, features of pseudo-laminar phases such as nonlinear correlations of increments of the laminar phases can be destroyed even by small noise strengths, where the influence of the noise is hardly visible.
	Since the concept of pseudo-laminar chaos is based on the general phenomenon of on-off intermittency and is not limited to a periodic driving, our results are not restricted to a specific system and therefore they are relevant for a large variety of problems, where the time derivative of an observable is an on-off intermittent signal.

	\begin{acknowledgments}
	    D.M-B. and G.R. gratefully acknowledge funding by the Deutsche Forschungsgemeinschaft (DFG, German Research Foundation) - 456546951. R.V. was supported by Australian Research Council (ARC) Discovery Project DP200100834 during the course of the work.
    \end{acknowledgments}

    \appendix

    \section{Pseudo-laminar chaos in switched systems}
    \label{sec:app_switched}
    
\begin{figure}[h]
		\includegraphics[width=0.48\textwidth]{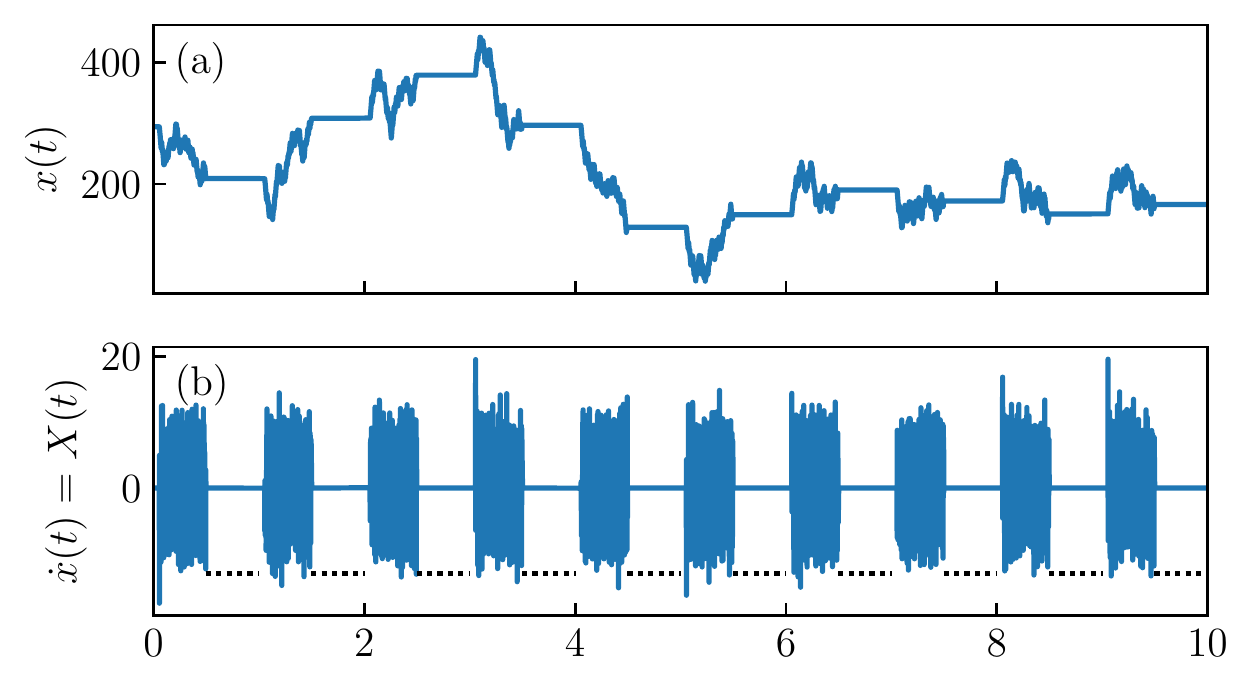}
		\caption{Time series of pseudo-laminar chaos in the switched system.
			In (a) pseudo-laminar chaos generated from Eq.~\eqref{eq:lorenz} is shown, where the parameters were chosen as in Fig.~\ref{fig:traj}(a).
			The sinusoidal variation of parameter $r=r(t)$ was replaced by a piecewise constant function, which equals $A_r$ if the equilibrium $(X,Y,Z)=(0,0,0)$ of the system in Fig.~\ref{fig:traj} is unstable (no dashed black line) and $-A_r$ if it is stable (black dashed line visible).
			The corresponding derivative of the time series are plotted in (b).
			For a sufficient relaxation of the transient dynamics, we begin plotting of the time series after $100$ periods of the time-varying parameter $r(t)$.
		}
		\label{fig:traj_app}
	\end{figure}

\begin{figure}[h]
		\includegraphics[width=0.48\textwidth]{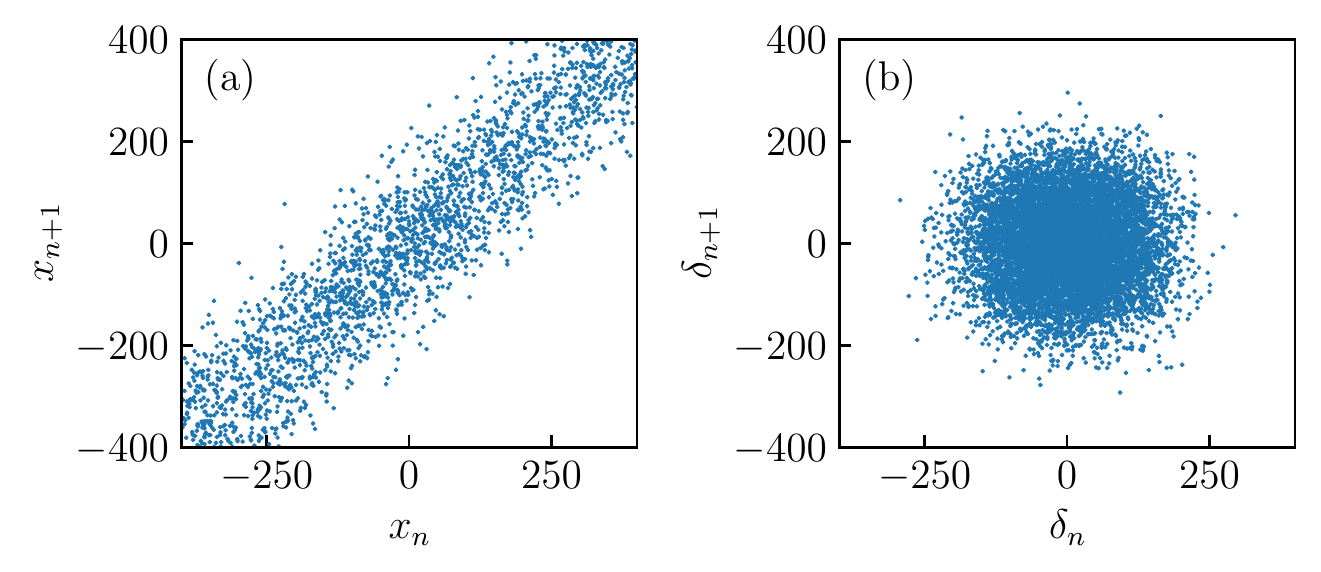}
		\caption{Return maps of the levels $x_n$ of the pseudo-laminar phases and of the increments $\delta_n = x_{n+1} - x_n$.
			The parameters were chosen as in Fig.~\ref{fig:traj_app}.
		}
		\label{fig:retmaps_app}
	\end{figure}

    In this section we consider Eq.~\eqref{eq:lorenz}, where the sinusoidal variation $r=r(t)=A_r \sin\left( \frac{2\pi}{T} t \right)$ is replaced by the piecewise constant function
    \begin{equation}
        \label{eq:rpiecewise}
        r(t)=
        \begin{cases}
            +A_r, & \text{if } A_r \sin\left( \frac{2\pi}{T} t \right) > 1 \\
            -A_r, & \text{else}.
        \end{cases}
    \end{equation}
    Equation~\eqref{eq:rpiecewise} preserves the durations of the pseudo-laminar phases and of the chaotic bursts observed for the sinusoidal parameter variation since the equilibrium $(X,Y,Z)=(0,0,0)$ of Eq.~\eqref{eq:lorenz} is stable for $r<1$ and unstable for $r>1$.
    Choosing the same parameter values as in Fig.~\ref{fig:traj}(a) and (b) we obtain qualitatively similar time series as shown in Fig.~\ref{fig:traj_app}.
    Comparing the return maps of the system with sinusoidal parameter variation shown in Fig.~\ref{fig:retmaps}(a) and (b) with the return maps of the switched system shown in Fig.~\ref{fig:retmaps_app} leads to the conclusion that the basic mechanisms of pseudo-laminar chaos can already be understood in a simpler periodically switched system, where well-defined chaotic and attractive fixed point dynamics alternate.

    Switched systems were extensively studied in \cite{liberzon_switching_2003,di_bernardo_piecewise-smooth_2008} and appear, for instance, in the implementation of chaotic logical circuits, where a nonlinear system is driven by a piecewise constant input signal, which carries the binary data to be processed.
    Results on logical circuits implemented by a thresholded chaotic Chua system can be found in \cite{rahman_qualitative_2018}.
    If the NOR gate considered there is fed with two identical periodic bit sequences, the output voltage of the threshold control unit shows periodically driven on-off intermittency and thus its integral shows pseudo-laminar chaos if the off state is normalized to zero.
    
    \section{Quantifying nonlinear correlations for the detection of laminar chaotic diffusion}
    \label{sec:app_corr}

    In order to distinguish laminar chaotic diffusion from pseudo-laminar chaos as done in Sec.~\ref{sec:lamchaostests} one has to check whether the dynamics of the levels $x_n$ of the nearly constant phases is governed by a one-dimensional map
    \begin{equation}
    	\label{eq:map}
    	x_{n+p} = f(x_n) = x_n + g(x_n).
    \end{equation}
    While this is clearly fulfilled for true laminar chaos as shown in Fig.~\ref{fig:retmaps}(e) and Fig.~\ref{fig:retmapsnoise}(e), for pseudo-laminar chaos, for all $p>0$, the points $(x_{n},x_{n+p})$ are randomly distributed around the bisectrix as shown in Figs.~\ref{fig:retmaps}(a,c) and Figs.~\ref{fig:retmapsnoise}(a,c).
    Interpreting $x_{n+p}$ and $x_n$ as random variables, the distinction between these cases leads to the problem of detecting and quantifying nonlinear correlations or stochastic dependence, which is a nontrivial problem \cite{tjostheim_statistical_2022}.
    In the following, we introduce a simple method for such a quantification and apply it to our numerical data for pseudo-laminar and true laminar chaos.
    The results are compared to the \emph{maximal information coefficient (MIC)} introduced in \cite{reshef_detecting_2011}.
    We find a qualitative agreement, which indicates that both methods are suitable for the distinction between pseudo-laminar and true laminar chaos.
    
    \begin{table}[h]
    	\caption{Comparison of $1-Q$ and the maximal information coefficient (MIC) for the return maps of the levels $x_n$ and of the increments $\delta_n = x_{n+1}-x_n$ shown in Fig.~\ref{fig:retmaps} and Fig.~\ref{fig:retmapsnoise}, for the systems without and with noise, respectively.
    	Values that are significantly larger than zero (bold) indicate that there are (nonlinear) correlations between subsequent levels or increments.
    	The sensitivity to noise of possible correlations in pseudo-laminar chaos is seen in the change of the coefficients from 1.00 for the increments in Fig.~\ref{fig:retmaps}(c) to a near-zero value after adding dynamical noise, Fig.~\ref{fig:retmapsnoise}(c).
    	}
    	\label{tab:corr}
    	\centering
    	\vspace{1mm}
    	\begin{tabular}{c|c|c|c|c}
    		& \multicolumn{2}{c|}{$1-Q$} &  \multicolumn{2}{c}{MIC}  \\
    		
    		parameters & levels  & increments  & levels  & increments  \\
    		\hline
    		Fig.~\ref{fig:retmaps}(a) & -0.00135 & -0.00142 & 0.062 & 0.063 \\
    		
    		Fig.~\ref{fig:retmaps}(c) & 0.00079 & \textbf{1.00} & 0.062 & \textbf{1.00} \\
    		
    		Fig.~\ref{fig:retmaps}(e) & \textbf{1.00} & \textbf{0.52} & \textbf{1.00}  & \textbf{0.66} \\
    		
    		Fig.~\ref{fig:retmapsnoise}(a) & 0.00232 & 0.0115 & 0.060  & 0.059 \\
    		
    		Fig.~\ref{fig:retmapsnoise}(c) & -0.0083 & 0.0051 & 0.060  & 0.062  \\
    		
    		Fig.~\ref{fig:retmapsnoise}(e) & \textbf{0.99} & \textbf{0.48} & \textbf{0.86}  & \textbf{0.62} \\
    	\end{tabular}
    \end{table}
    
    Since the intensity levels $x_n$ show diffusive dynamics, there is always a strong linear correlation between subsequent levels $x_n$ and $x_{n+p}$, which overlay the relevant nonlinear correlations induced by $g$ in Eq.~\eqref{eq:map} and leads to wrong results if the algorithms used here are applied directly to the return map generated by $(x_n,x_{n+p})$.
    Instead we consider the points $(x_n,x_{n+p}-x_n)=(x_n,\delta_n)$, which resemble the graph of the function $g$ in Eq.~\eqref{eq:map} for true laminar chaos, whereas, for pseudo-laminar chaos the $\delta_n$ fluctuate around zero without correlations with $x_n$.
    To obtain comparable results for the quantification of the correlations, we require that the used quantity converges, for an increasing number of samples, to $0$ if and only if the two considered variables are stochastically independent and that it converges to $1$, if there is an exact functional relationship, such as Eq.~\eqref{eq:map}, between these variables, provided that this function is not constant.
    Our approach follows from the same idea as the CANOVA method, which can be found in \cite{wang_efficient_2015}:
    Assuming that the function $g$ is continuous, $|g(x)-g(x')|$ is small for small $|x-x'|$, i.e., if the arguments $x$ and $x'$ are close.
    If the points $(x_n,\delta_n)$ resemble the graph of the function $g$, it follows that $|\delta_{n'} - \delta_n|$ is small if $x_n$ and $x_{n'}$ are close.
    A normalized quantity $Q$ that reflects this property is obtained by relating the variance of $\delta_n-\delta_{n'}$ for close $x_n$ and $x_{n'}$ to the overall variance of the $\delta_n$.
    To compute the quantity $Q$, we first sort the sequence of points $(x_n, \delta_n)$ with respect to the first component $x_n$ and obtain a sequence of points $(v_m,w_m)$, with $v_m = x_{\pi^{-1}(m)}$ and $w_m = \delta_{\pi^{-1}(m)}$, where the indices $m$ are given by a permutation $m=\pi(n)$ of the indices $n$ such that we have $v_{m} < v_{m+1}$ for all $m$.
    The quantity $Q$ is defined by
    \begin{eqnarray}
    	\label{eq:Q}
    	Q = \frac{\text{Var}(w_{m+1}-w_m)}{2\, \text{Var}(w_m)},
    \end{eqnarray}
    where $\text{Var}(w_m)$ is the sample variance of all values $w_m$.
    If the $x_n$ fulfill Eq.~\eqref{eq:map}, which implies $w_m = g(v_m)$, we have $w_{m+1}-w_m = g(v_{m+1})-g(v_m)$.
    Assuming that the $x_n$ densely fill an interval (or a union of intervals), we have, in the limit of an infinite number of samples, $v_{m+1} - v_m \to 0$ for (almost) all $m$.
    It follows that the numerator converges to zero in this limit and we have $Q\to 0$, provided that $g(x)$ is not constant, which would lead to a vanishing denominator.
    If $w_m=\delta_n$ is discrete white noise, we have $\text{Var}(w_{m+1}-w_m) \to 2\, \text{Var}(w_m)$ in the limit of an infinite number of samples so that we have $Q=1$ in this case.
    Considering $1-Q$ instead of $Q$ gives values close to $1$ or $0$ for true laminar or pseudo-laminar chaos, respectively. 
    An implementation of the algorithm can be found in Listing~\ref{ls:Q}.
    In Tab.~\ref{tab:corr}, the quantity $1-Q$ is shown for the return maps of the levels $x_n$ and the increments $\delta_n$ of the levels of the (pseudo-)laminar phases visualized in Fig.~\ref{fig:retmaps} and Fig.~\ref{fig:retmapsnoise}, where the algorithm was applied to $10^4$ points $(x_n,\delta_n)$ and $(\delta_n,\delta_{n+1})$, respectively.
    
    \begin{lstlisting}[language=Mathematica, otherkeywords={SortBy}, caption={Mathematica implementation of the algorithm for the computation of $Q$. For computing $Q$ of the return map of the levels $x_n$ or the increments $\delta_n$, the variable \texttt{data} must contain the list of all points $(x_n,\delta_n)$ or $(\delta_n,\delta_{n+p})$, respectively.}, label=ls:Q]
w = SortBy[data, First][[All, 2]];
Q = Variance[
        w[[2 ;; -1]] - w[[1 ;; -2]]
    ]/(2*Variance[w]);
    \end{lstlisting}
    
    The results are compared to the \emph{maximal information coefficient (MIC)} introduced in \cite{reshef_detecting_2011}, which converges, as the quantity $1-Q$, to $0$ for stochastically independent variables and it converges to $1$ if the variables are connected by a nonconstant function.
    The authors there also demonstrate that different functions lead to similar values of the MIC if noise with the same noise strength is added.
    So it should be an appropriate quantity for detecting nonlinear correlations during the test for laminar chaos.
    To analyze correlations in the return maps of the levels $x_n$ or the increments $\delta_n$, the algorithm is applied to sets of points $(x_n,\delta_n)$ or $(\delta_n,\delta_{n+p})$, respectively.
    The MIC values were computed from $10^4$ data points with $p=1$ using \emph{minepy} \cite{albanese_minerva_2013}.
    
    For the considered parameters, both methods are able to distinguish pseudo-laminar chaos from true laminar chaos even in the case shown in Fig.~\ref{fig:retmaps}(c), where the return map of the levels of the pseudo-laminar phases basically consists of two lines.

	\bibliography{pseudolaminarchaos}
	
\end{document}